\documentclass[10pt,letterpaper]{article}
\usepackage{opex3}
\usepackage{color}
\usepackage{cite}

\begin{document} 
\title{Quantum frequency conversion of quantum
memory compatible photons to telecommunication wavelengths}
\author{Xavier Fernandez-Gonzalvo,$^{1,2}$ Giacomo
Corrielli,$^{1,3}$ Boris Albrecht,$^{1}$ \\Marcel.li
Grimau,$^{1,4}$ Matteo Cristiani,$^{1*}$ and Hugues de
Riedmatten$^{1,5}$} \address{$^1 $ ICFO-Institut de Ciencies
Fotoniques, Mediterranean Technology Park, 08860 Castelldefels (Barcelona), Spain.\\
$^2 $ Current address: Department of Physics,
University of Otago, 730 Cumberland Street, 9016 Dunedin, New Zealand \\
$^3 $ Current address: Politecnico di Milano, Dipartimento di
Fisica - Piazza Leonardo da Vinci, 32, I-20133 Milano, Italy\\ $^4
$ Current address: Institute for Quantum Information Science and
Department of Physics and Astronomy, University of Calgary,
Calgary, Alberta T2N 1N4, Canada\\ $^5 $ ICREA-Instituci\' o
Catalana de Recerca i Estudis Avan\c cats, 08015 Barcelona, Spain
} \email{$^*$matteo.cristiani@icfo.es}

\begin{abstract}
We report an experiment demonstrating quantum frequency conversion
of weak light pulses compatible with atomic quantum memories to
telecommunication wavelengths. We use a PPLN nonlinear waveguide
to convert weak coherent states at the single photon level with a
duration of $30\,\mathrm{ns}$ from a wavelength of
$780\,\mathrm{nm}$ to $1552\,\mathrm{nm}$. We measure a maximal
waveguide internal (external) conversion efficiency $\eta_{int}=
0.41$ ($\eta_{ext}=0.25$), and we show that the signal to noise
ratio ($SNR$) is good enough to reduce the input photon number
below 1. In addition, we show that the noise generated by the pump
beam in the crystal is proportional to the spectral bandwidth of
the device, suggesting that narrower filtering could significantly
increase the $SNR$. Finally, we demonstrate that the quantum
frequency converter can operate in the quantum regime by
converting a time-bin qubit and measuring the qubit fidelity after
conversion.
\end{abstract}



\bibliographystyle{osajnl}

\section{Introduction}
Quantum repeaters \cite{Briegel1998,Duan2001,Sangouard2011} have
been proposed as a potential solution to overcome the problem of
exponential loss in optical fibers, that prevents the direct
distribution of quantum information beyond a few hundred
kilometers. Quantum repeaters rely on heralded entanglement
between remote quantum memories (QMs) \cite{Duan2001,
Sangouard2011}. In absence of a heralding mechanism for the
absorption of a photon in the QM, heralded entanglement is usually
obtained by a measurement induced mechanism (e.g. a Bell state
measurement) with light emitted by the remote QMs
\cite{Duan2001,Chou2005,Yuan2008}. Despite initial work towards
QMs operating in the telecom band
\cite{Lauritzen2010,Lauritzen2011} most QMs operate in a
wavelength range where the loss in optical fibers is significant.
Hence a quantum interface allowing to connect these QMs to optical
fibers by converting the emitted photons to telecom wavelengths is
needed for almost all applications in the context of quantum
communication \cite{Sangouard2011, Shahriar2012}.

Quantum frequency conversion has attracted a lot of attention
recently. Several nonlinear processes have been used, including
frequency upconversion using sum frequency generation
\cite{Huang1992, Tanzilli2005, Rakher2010, Ates2012}, frequency
down conversion using difference frequency generation (DFG)
\cite{Ding2010, Curtz2010, Takesue2010, Zaske2011, Zaske2012,
DeGreve2012, Ikuta2011} and four wave-mixing \cite{Radnaev2010,
McGuinness2010}. DFG enables the conversion of visible or near
infrared light to telecommunications wavelengths and is thus
ideally suited for quantum repeater applications. The first proof
of principle experiments of DFG at single photon level were
reported in \cite{Curtz2010, Takesue2010}. Progress has been fast
since this first demonstration, including high efficiency
conversion \cite{Zaske2011} and conversion of non classical light
emitted by single solid state emitters \cite{Zaske2012,
DeGreve2012} and by parametric down conversion
sources\cite{Ikuta2011}. However, most of the experiments done so
far with DFG in nonlinear materials were done with very short
pulses, not compatible with the narrow bandwidth of current
emissive quantum memories, which typically emit transform limited
single photon pulses with a few tens of ns duration
\cite{Felinto2006, Sangouard2011}. The realization of a
narrowband, quantum memory compatible quantum frequency converter
(QFC) is significantly more demanding in terms of noise
suppression, as the duration of the converted photons is
significantly longer than the single photon emitted by solid state
emitters \cite{Ates2012, Zaske2012, DeGreve2012} and spontaneous
down conversion sources \cite{Tanzilli2005, Ikuta2011}.

A proof of principle experiment of quantum frequency conversion
from a Rubidium quantum memory to $1367\,\mathrm{nm}$ has been
reported recently, using four wave mixing in a cold and  extremely
dense atomic ensemble \cite{Radnaev2010}. This technique, while
efficient and noise free, requires a complex experimental setup
and is restricted to wavelengths close to atomic transitions. In
contrast, quantum frequency conversion based on nonlinear
processes in $\chi^{(2)}$ or $\chi^{(3)}$ materials is more
versatile, since it allows an almost free choice of the signal and
converted wavelengths. This will be important in view of
connecting quantum systems of different kinds. In addition the
experimental setup is significantly simpler and can be integrated
using waveguide technology.

In this paper, we report for the first time to our knowledge a
photonic solid state quantum interface enabling the quantum
frequency conversion of photons compatible with a Rubidium QM to
telecommunication wavelengths. We use DFG in a periodically poled
lithium niobate (PPLN) waveguide with a strong pump at
$1569\,\mathrm{nm}$, to convert weak atom resonant coherent pulses
of $30\,\mathrm{ns}$ duration at the single photon level from
$780\,\mathrm{nm}$ to $1552\,\mathrm{nm}$. We show that the signal
to noise ratio ($SNR$) is good enough to reduce the average input
photon number $\mu_{in}$ below $1$. In addition, we show that the
noise decreases linearly with the spectral bandwidth of the
device, suggesting that a narrower filtering should allow us to
decrease the minimum input photon number required to achieve a
$SNR > 1$. Finally, we measure the fidelity of the conversion
process by using time-bin input qubits, and show that our device
can operate with conditional  fidelities higher than classical
{\it measure and prepare} strategies taking into account the
statistics of the input coherent states and the finite device
efficiency.

The paper is organized as follows. The experimental setup is
described in section \ref{sec:expset}. Section \ref{sec:devchar}
is devoted to the characterization of the frequency conversion
device, including a study of the efficiency of the process and of
the related noise. In section \ref{sec:coh}, we report the
experiments related to the phase preservation of the DFG process
and finally in section \ref{sec:dis} we discuss in more detail the
performances of our device, and possible improvements.

\begin{figure}[htbp]
\begin{center}
\includegraphics[width=0.9\textwidth]{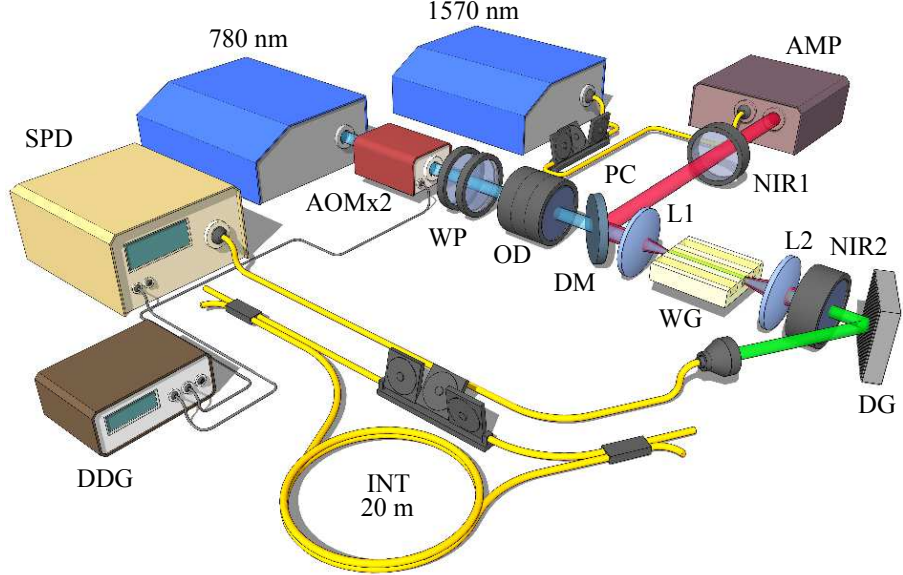}
\caption{\label{fig:setup}Schematic view of the experimental
setup. The pump laser is a diode at $1570\,\mathrm{nm}$. PC:
polarization controller. AMP: fiber amplifier. NIR1: band-pass
filter to clean the ASE. The input light is obtained by a
diode--tapered amplifier system at $780\,\mathrm{nm}$. AOMx2:
acousto-optic modulator in double passage. WP: half- and quarter-
waveplates. OD: neutral density filters. The pump and input beams
are overlapped on a dichroic mirror DM. L1: in-coupling lens. WG:
non-linear waveguide. L2: out-coupling lens. NIR2: band-pass
filter centered at $1552\,\mathrm{nm}$ plus long-pass filter with
cut-off at $1450\,\mathrm{nm}$. DG: diffraction grating. SPD:
single photon detector. DDG: digital delay generator. INT: fiber
interferometer.}
\end{center}
\end{figure}

\section{Experimental setup}
\label{sec:expset} In the following section we will describe the
main components of our frequency conversion unit. The experimental
setup is schematically depicted in Fig. \ref{fig:setup}.

\subsection{Laser sources}
The pump light at $\lambda_p = 1569.4\,\mathrm{nm}$ is provided by
an external cavity diode laser (ECDL) amplified by a continuous
wave Erbium doped fiber amplifier. In order to remove the
amplified spontaneous emission (ASE) we use two bandpass filters
centered at $1570\,\mathrm{nm}$ with a bandwidth of
$9\,\mathrm{nm}$ (NIR1, Semrock). The peak transmission of the
filters at $1569\,\mathrm{nm}$ is $>98\,\%$, while around
$1552\,\mathrm{nm}$ it is $<10^{-5}$. The ASE is then suppressed
by $>100\,\mathrm{dB}$ at the converted wavelength.

The input light at $\lambda_{in} = 780.24\,\mathrm{nm}$ is
obtained from a commercial system formed by an ECDL, followed by a
tapered amplifier (Toptica TA pro). The laser is tuned to be
resonant with the $\mathrm{D_2}$ line of $\mathrm{{}^{87}Rb}$. In
order to simulate photons obtained from a rubidium based quantum
memory, about $20\,\mathrm{mW}$ of the total available power is
sent through an acousto-optic modulator (AOMx2) arranged in double
passage configuration. Pulses are created with a full width half
maximum (FWHM) of $30\,\mathrm{ns}$, compatible with the
$26.2\,\mathrm{ns}$ life time of the $\mathrm{5{}^2P_{3/2}}$
excited state of $\mathrm{{}^{87}Rb}$ and with durations of single
photons emitted by quantum memories based on cold atomic ensembles
\cite{Felinto2006}. A digital delay generator (DDG) is used to
control the AOM and generate pulses at a repetition rate of
$1\,\mathrm{MHz}$. The input light is then fiber coupled (fiber
not shown in Fig. \ref{fig:setup}) to clean its spatial mode. At
the fiber output, a set of calibrated neutral density filters (OD)
attenuate the laser intensity. The total optical depth given by
the filters is measured to be $16.6(1)$. The mean photon number
per pulse before the waveguide $\mu_{in}$ is then spanned from
$0.3$ to $25$ by fine tuning of the laser power.

\subsection{PPLN waveguide}
\label{sec:coup} The DFG process is based on Type $0$ quasi
phase-matching in a $L = 3\,\mathrm{cm}$ periodically poled
lithium niobate waveguide ($5\,\%$ $\mathrm{MgO}$ doped congruent
PPLN, AR-coated for $780$ and $1552$--$1569\,\mathrm{nm}$, HC
Photonics Corp.). The waveguide cross section of
$7.08\times11.35\,\mathrm{{\mu m}^2}$ ensures single mode
operation at $1552$--$1569\,\mathrm{nm}$, while for the input
light at $780\,\mathrm{nm}$ the propagation is multi-mode. The
transmission losses specified by the manufacturer are
$0.7\,\mathrm{dB/cm}$ ($0.35\,\mathrm{dB/cm}$) at
$780\,\mathrm{nm}$ ($1552$ -- $1569\,\mathrm{nm}$). The waveguide
is mounted on a copper holder, which is temperature stabilized by
means of a Peltier unit used in combination with a home-made
controller. The chip temperature is maintained around
$320\,\mathrm{K}$, to ensure quasi-phase matching and thus
maximizing the conversion efficiency.

The pump light polarization is adjusted to be oriented along the
proper crystal axis by means of a polarization controller (PC)
placed between the seed laser and the amplifier. For the input
light the polarization is adjusted with a combination of a half-
and a quarter-waveplate (WP). The input and pump beams are
overlapped on a dichroic mirror (DM) and coupled to the waveguide
by means of an aspheric lens (L1) with focal length
$f=15.9\,\mathrm{mm}$ and numerical aperture $NA=0.16$. The
in-coupling lens is mounted on an XYZ flexure stage for fine
adjustment of the waveguide coupling. The converted light is then
out-coupled by means of a second aspheric lens (L2) with the same
focal length and numerical aperture as L1. The transmission at
$1570\,\mathrm{nm}$ is measured to be higher than $98\,\%$, while
only $80\,\%$ of the input light is transmitted.

\begin{table}
\begin{center}
\caption{Summary of the losses at the different elements.}
\label{tab:loss}
\begin{tabular}{|l|c|c|}
\hline Transmission  & $\lambda_{in} = 780\,\mathrm{nm}$ & $\lambda_p = 1569\,\mathrm{nm}$\\
\hline
Input lens (L1) & 0.99 & $0.66$  \\
Waveguide (WG) coupling & 0.61 & 0.58 \\
Waveguide (WG) propagation & 0.61 & $0.78$ \\
Output lens (L2) & 0.80 & 0.98 \\
\hline
Total  & 0.29& 0.30 \\
\hline
\end{tabular}
\end{center}
\end{table}

The transmission of the pump (input) light through the converter
is measured to be $30\,\%$ ($29\,\%$). These values account for
the coupling efficiency  in the waveguide and the losses
experienced during the propagation through the crystal as well as
the coupling lenses. Using the transmission values specified
above, we infer that the coupling efficiency of the pump (input)
in the waveguide is $58\,\%$ ($61\,\%$). Table \ref{tab:loss}
summarizes the various losses in the waveguide and associated
optics.

Note that the pump light attenuation does not present a
significant problem, since the total available power is much
higher than the value required to operate the experiment. On the
other hand, the waveguide transmission at $\lambda_{in}$ is a
crucial parameter, since it affects the device efficiency and a
lower transmission reduces the $SNR$ for the converted photons. It
is also important to note that the transmission at
$780\,\mathrm{nm}$ has been measured after optimizing the DFG
signal with classical light. The fact that the waveguide is
multimode at $780\,\mathrm{nm}$ and that the input and the pump
modes have different mode diameters indicates that the maximum
conversion efficiency does not necessarily corresponds to the
maximum input coupling. The waveguide coupling at the input
wavelength could be increased by using ridge waveguides for which
it has been shown that coupling efficiencies up to $90\,\%$ can be
achieved \cite{Zaske2011}.

\subsection{Filtering stage}
\label{sec:filter} A fundamental part of our setup is the
filtering stage used to isolate the converted signal at
$\lambda_{out} = 1552\,\mathrm{nm}$ from the noise. In this type
of quantum frequency conversion experiments two main contributions
to the noise in the vicinity of $\lambda_{out}$ can arise: leakage
of the strong pump at $1569\,\mathrm{nm}$ and spontaneous Raman
scattering \cite{Langrock2005, Zaske2011}. In our case spontaneous
parametric down conversion (SPDC) of the pump photons
\cite{Pelc2010} can not contribute to the noise, since
$\lambda_{out} < \lambda_p$. It has also to be noted that light at
$785\,\mathrm{nm}$ could be created by weakly phase matched second
harmonic generation (SHG) of the strong pump.

In order to remove the unconverted input photons and the light
created by SHG of the pump we use a long pass filter with cut-off
wavelength of $1450\,\mathrm{nm}$  (Thorlabs), placed right after
the waveguide. Its transmission at $1552\,\mathrm{nm}$ is
$\sim85\,\%$, while its optical depth in the $700$ -
$800\,\mathrm{nm}$ range is $\sim 5$. The residual pump light is
then blocked by means of two band-pass filters centered at
$1552\,\mathrm{nm}$ with a bandwidth of $7\,\mathrm{nm}$ (Semrock,
transmission $\sim93\,\%$).

A diffraction grating ($600$ lines per $\mathrm{mm}$,
$1.6\,\mathrm{\mu m}$ blaze, Thorlabs) is then used to map the
frequency spectrum of the converted light into angular dispersion.
The incidence angle of $3\,\mathrm{deg}$ has been chosen to
increase the spectral resolution while maintaining a sufficient
diffraction efficiency. The light diffracted in the first order is
then coupled to a single mode fiber, thus selecting only a narrow
spectral region. This method offers two advantages. On the one
hand, narrower transmission bandwidths are achievable. On the
other hand, this solution allows us to vary the filtering width up
to a certain extent. This is achieved by changing the focal
position of the out-coupling lens, thus modifying the aspect ratio
of the converted beam at the position of the fiber coupler. When
the beam is elongated in a direction perpendicular to the grating
dispersion, the minimum filter bandwidth is achieved. When the
aspect ratio is inverted, the filter resolution is reduced. This
allows us to span the transmission bandwidth from $0.65
\,\mathrm{nm}$ ($80\,\mathrm{GHz}$) to $2.3\,\mathrm{nm}$
($287\,\mathrm{GHz}$).

The total transmission of the filtering stage at $\lambda_{out}
=1552\,\mathrm{nm}$ is about $26\,\%$. This number accounts for
the single mode fiber coupling ($\sim50\,\%$), the diffraction
grating efficiency ($70\,\%$), and the transmission of the
long-pass and band-pass filters ($74\,\%$).

The converted photons are finally detected by means of a single
photon detector (SPD, id201, idQuantique) operated in gated mode,
with an overall detection efficiency of $7\,\%$, which includes
the detector efficiency ($10\,\%$) and a fiber to fiber connection
($70\,\%$). The dark count probability is about
$10^{-5}\,\mathrm{ns^{-1}}$. In order to avoid afterpulses we
adopt a dead time of $20\,\mathrm{\mu s}$. The photon counter is
triggered by the DDG in order to synchronize the detection window
with the photon arrival time.

\section{Device characterization}
\label{sec:devchar} In this section we will describe different
measurements intended to characterize the device operation and
efficiency.
\subsection{Pulse width and detection window}
\label{sec:FWHM} 
\begin{figure}[htbp]
\begin{center}
\begin{tabular}{cc}
\includegraphics[width=0.45\textwidth]{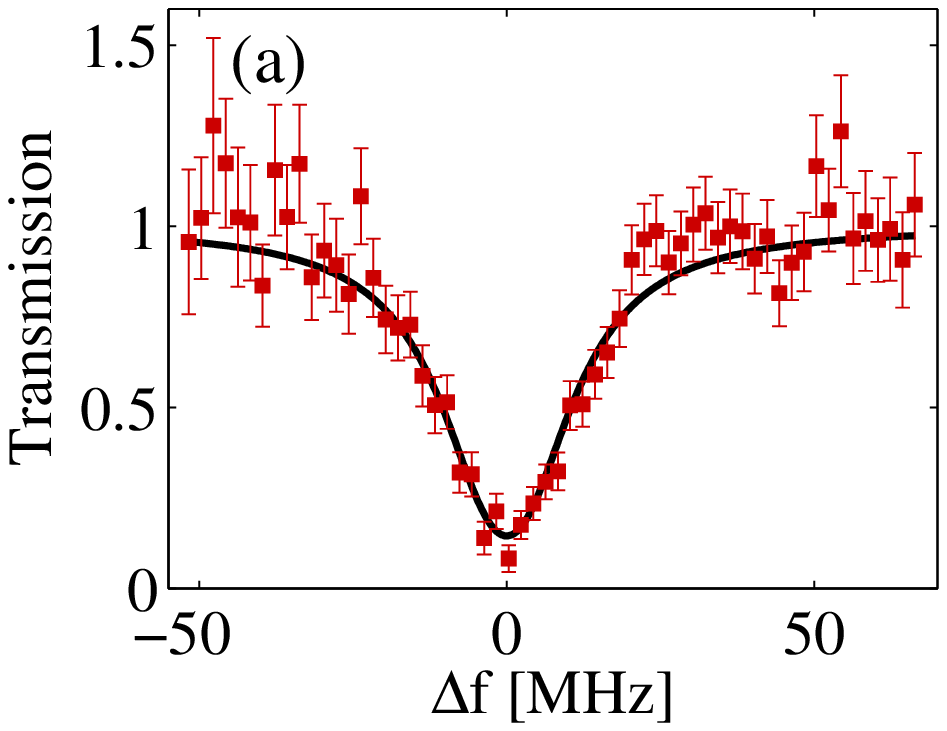} &
\includegraphics[width=0.45\textwidth]{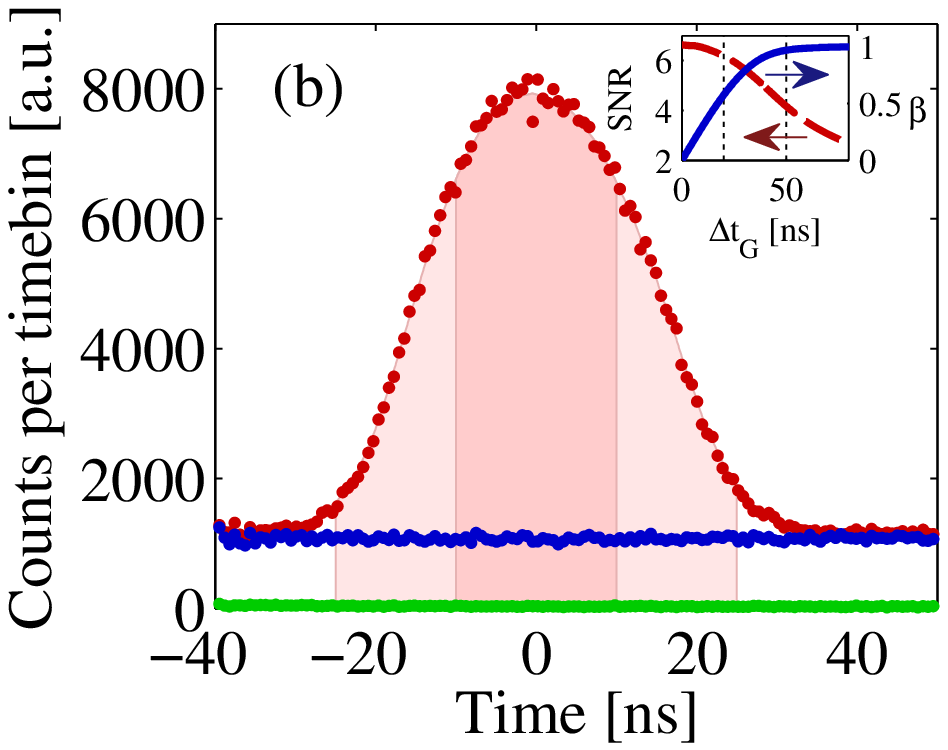}
\end{tabular}
\caption{\label{fig:pulse} {\bf (a)} Transmission of weak input
pulses through a cloud of cold $\mathrm{{}^{87}Rb}$ atoms as a
function of detuning. The probe light at $780\,\mathrm{nm}$ is
converted to $1552\,\mathrm{nm}$ after interacting with the atomic
cloud  and then detected with the SPD. {\bf (b)} Temporal shape of
the converted photons. The time-bin was $0.64\,\mathrm{ns}$. The
shaded areas represent the detector gates of $20$ and
$50\,\mathrm{ns}$. Inset: signal to noise ratio (left axis) and
$\beta$ factor (right axis) as a function of the detection window.
The plot is obtained using the data points shown in the main
figure (see text for details).}
\end{center}
\end{figure}

As stated in the previous section, the input
pulses are characterized by a wavelength of $\lambda_{in} =
780\,\mathrm{nm}$ and a FWHM of $30\,\mathrm{ns}$ to mimic photons
obtained from DLCZ quantum memories \cite{Felinto2006}. In order
to verify their compatibility with Rubidium based atomic
ensembles, we send weak input pulses to a cold cloud of
$\mathrm{{}^{87}Rb}$ atoms confined in a magneto-optical trap.
This is part of another setup present in the same laboratory and
used to implement an ensemble based QM. After interacting with the
atomic ensemble, the light pulses are sent to the QFC and the
corresponding $1552\,\mathrm{nm}$ converted photons are detected.
The transmission of the input photons through the atomic cloud as
a function of the input frequency is plotted in Fig.
\ref{fig:pulse}(a). A typical absorption profile can be observed.
The efficient absorption confirms that the input pulses interact
strongly with the atoms. A fit to the experimental data gives a
value for the full width at half maximum of the absorption peak of
$16(1)\,\mathrm{MHz}$. This value is close to the convolution
between the natural linewidth of the transition
($6\,\mathrm{MHz}$) and the bandwidth of the close to Gaussian
input photons ($0.44/(30\,\mathrm{ns})=14\,\mathrm{MHz}$).

We then measure the photon shape after the QFC, in order to verify
that it is preserved during the conversion process. We prepare
input pulses with a mean photon number $\mu_{in}=5$ and send them
to the converter. By means of the DDG, we trigger the photon
counter such that the detection window of $100\,\mathrm{ns}$ is
centered around the photon arrival time. The counts given by the
single photon detector are then acquired by a time stamping card
(Signadyne) together with the trigger signal. We finally perform a
start-stop measurement, using the trigger as start and a
$1552\,\mathrm{nm}$ detection as stop. The results are shown in
Fig. \ref{fig:pulse}(b). The detector dark counts per time-bin
($dc$) are obtained by blocking the waveguide input during the
measurement (green dots). The noise level ($n$, including dark
counts) is obtained in the presence of the pump light only,
blocking the $780\,\mathrm{nm}$ input (blue dots). The signal
($s$, including noise and dark counts) is plotted with red dots.

In a typical experiment, the detected counts are integrated over a
certain time window characterized by a width $\Delta t_G$. In the
following, we will refer to the total signal as $S=\int_{\Delta
t_G} s\,\mathrm{d}t$. Similar definitions are employed for the
total noise ($N$) and dark counts ($DC$). Two figures of merit
guide us in the choice of $\Delta t_G$: the detected signal
fraction $\beta$ and the signal to noise ratio $SNR$. $\beta$ is
defined as $(S-N)/(S'-N')$, where $S'$ ($N'$) is the signal
(noise) for $\Delta t_G \rightarrow \infty$. The signal to noise
ratio with dark counts subtracted is defined as $SNR = (S - N) /
(N - DC)$. On the one hand, $\beta$ increases by extending the
detection region, reaching the asymptotic value of $1$ for $\Delta
t_G \rightarrow \infty$. On the other hand, as $\Delta t_G$
increases the $SNR$ drops to $0$, since the noise increases
linearly with $\Delta t_G$, while the signal fast saturates to
$S'$ for $\Delta t_G
> 30\,\mathrm{ns}$. This behavior is illustrated in the inset of
Fig. \ref{fig:pulse}(b).

The single photon detector we use allows only for certain
detection window widths. The values closer to the pulse FWHM used
in the experiments are $20$ and $50\,\mathrm{ns}$, represented as
red shaded areas in Fig. \ref{fig:pulse}(b). While $\Delta t_G =
50\,\mathrm{ns}$ corresponds to the highest detected signal
fraction ($\beta=0.97$), $\Delta t_G = 20\,\mathrm{ns}$ offers the
highest $SNR$ with $\beta=0.57$. Unless otherwise specified, for
the measurement described in the following we adopt a
$20\,\mathrm{ns}$ detection window.

\subsection{Noise vs pump power}
\label{sec:N}The noise level $N$ introduced in the previous
section is related to the pump power $P_p$. We measured the noise
as a function of $P_p$ in absence of input signal, as shown in
Fig. \ref{fig:SNR}(a) , and found that it follows a linear
relation:
\begin{equation}
    \label{eq:N}
    N\left(P_p\right) = \alpha \cdot
    P_p + DC,
\end{equation}
This behavior suggests that the noise is originated either by pump
leakage or by spontaneous Raman scattering. A fit of the
experimental data gives $\alpha \sim 6\times10^{-6}/\mathrm{mW}$
for $\Delta t_G=20\,\mathrm{ns}$. This value expresses the noise
level for detected events. If we back propagate to the crystal by
correcting for optical losses (see section \ref{sec:eff}), we find
that the noise floor is $\alpha^\prime = 10^{-4}/\mathrm{mW}$ for
$\Delta t_G=20\,\mathrm{ns}$. It is important to notice that this
value is valid only for the filtering width of $\sim
0.7\,\mathrm{nm}$ ($85\,\mathrm{GHz}$) used for the measurements
discussed in the following.

The fact that the noise is linearly related to the pump power with
a slope which has been properly calibrated, allows us to consider
$N$ as an indirect measurement of $P_p$. This technique simplifies
the determination of the pump power and it has been used to obtain
the quantity plotted in the horizontal axis of Fig.
\ref{fig:SNR}(b), discussed in the following section.

\subsection{Conversion and total efficiencies}
\label{sec:eff} 
\begin{figure}[htbp]
\begin{center}
\begin{tabular}{cc}
\includegraphics[width=0.428\textwidth]{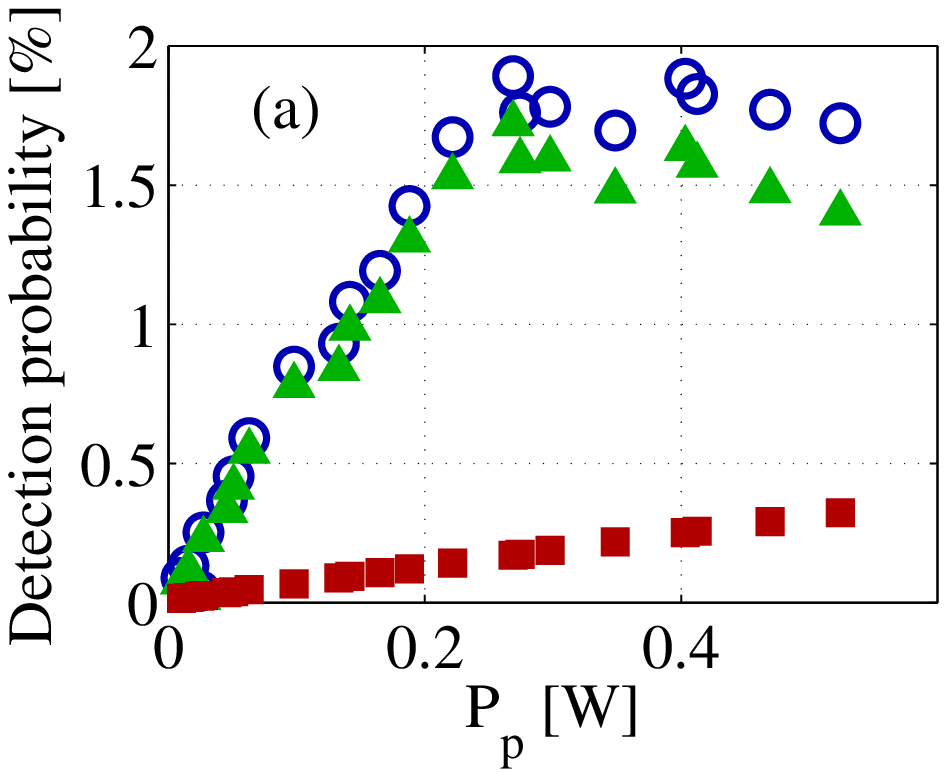} &
\includegraphics[width=0.472\textwidth]{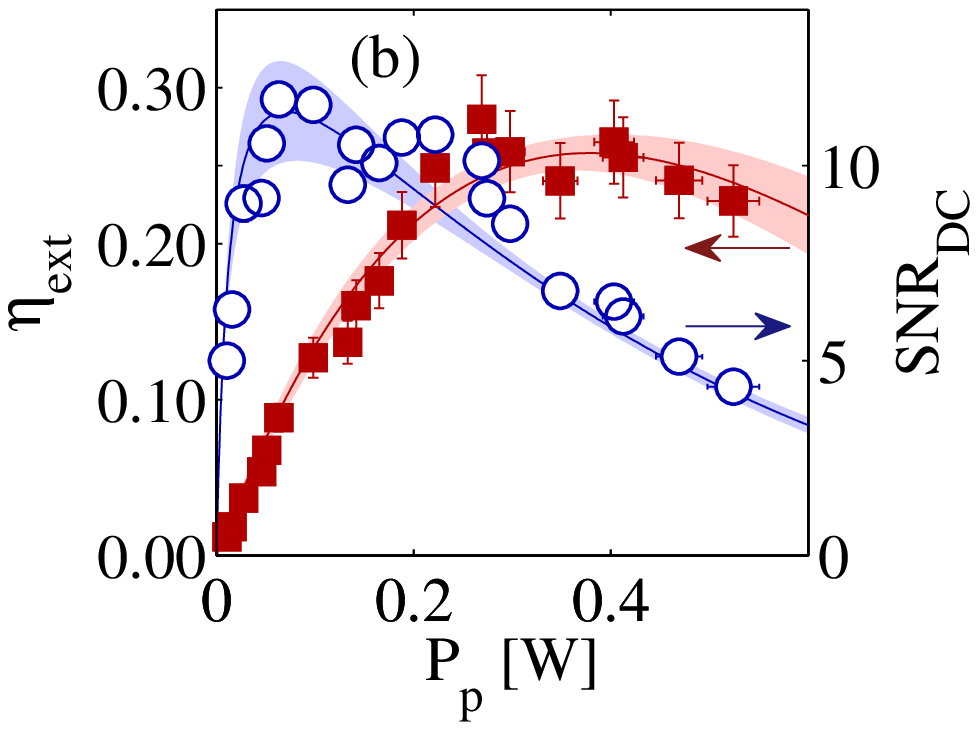}
\end{tabular}
\caption{\label{fig:SNR}{\bf (a)} Detection probabilities with
($p_S$, blue open circles) and without ($p_N$, red squares) input
signal ($\mu_{in}=6.1$)  as a function of the pump power $P_p$
measured after the waveguide. The green triangles correspond to
the pure signal ($p_S-p_N$). {\bf (b)} External conversion
efficiency ($\eta_{ext}$)  as a function of $P_p$ (red dots, left
axis). The continuous line is a fit using Eq. (\ref{eq:eta}). The
red shaded area represents the $95\,\%$ confidence interval of the
fit. We observe a normalized conversion efficiency of $\eta_n =
650 (70)\,\mathrm{\% / W}$. On the right axis we show the signal
to noise ratio. The experimental data (blue dots) are compared
with Eq. (\ref{eq:snr}), where all the parameters are determined
independently. The blue shaded area accounts for the errors
associated to the different quantities involved in Eq.
(\ref{eq:snr}).}
\end{center}
\end{figure}

In order to measure the total efficiency, we send
weak coherent pulses with $\mu_{in} = 6.1$ and we measure the
probability to detect a photon per gate with ($p_S$) and without
input pulse ($p_N$) as a function of the pump power, as shown in
Fig. \ref{fig:SNR}(a) \cite{Zaske2011}. The total efficiency,
including all losses, is then calculated by
$\eta_{tot}=(p_S-p_N)/\mu_{in}$.

We find a maximum total efficiency $\eta_{tot}^{M}$=$2.6 \times
10^{-3}$. By correcting for the detection efficiency ($\eta_{det}
=0.07\times \beta =0.04$), we can infer the maximal device
efficiency $\eta_{dev}^{M}= 0.066$, which can be interpreted as
the probability to find a photon in a single mode fiber at the
output of the conversion device (including spectral filtering)
with $\mu_{in}=1$. Finally, by correcting for the transmission of
the filtering system ($0.26$) we can compute the external
conversion efficiency of the waveguide, corresponding to
$\eta_{ext}=\mu_{out}/\mu_{in}$ where $\mu_{out}$ is the average
number of photons at $1552\,\mathrm{nm}$ at the output of the
waveguide ($\eta_{ext}^{M}=0.25$). $\eta_{ext}$ is limited by the
coupling into the waveguide ($\eta_c=0.61$), so the maximal
internal waveguide conversion efficiency is $\eta_{int}^{M}=0.41$.
The various efficiencies and losses are summarized in Table
\ref{tab:eff}.

\begin{table}
\begin{center}
\caption{Efficiency definitions and values. See text for details.}
\label{tab:eff}
\begin{tabular}{|l||c|c||c|c|}
\hline
Description & \multicolumn{2}{|c|}{Individual} & \multicolumn{2}{|c|}{Cumulative}\\
\hline
Coupling efficiency & $\eta_{c}$ & $0.61$ & $\eta_{int}^M$ & $0.41$ \\
Effective waveguide transmission & $\eta_{t}$ & $0.7$ & $\eta_{ext}^M$ & $0.25$ \\
Filter transmission & $\eta_{f}$ & $0.26$ & $\eta_{dev}^M$ & $0.066$ \\
Detection efficiency (incl. $\beta$)& $\eta_{d}$ & $0.04$ & $\eta_{tot}^M$ & $0.0026$\\
\hline
\end{tabular}
\end{center}
\end{table}

In Fig. \ref{fig:SNR}(b), the external efficiency is plotted as a
function of the pump power measured after the waveguide $P_p$ (red
dots, left axis). The continuous line is a fit of the experimental
data with the formula:
\begin{equation}
    \label{eq:eta}
    \eta_{ext} = f\left(P_p\right) =
    \eta_{ext}^M\sin^2\left( L \sqrt{P_p \cdot \eta_{n}} \right),
\end{equation}
where $\eta_{n}$ is the normalized efficiency and $L$ the
waveguide length. This formula is obtained by treating the pump as
a classical field and assuming that the converted photon input is
in the vacuum state \cite{Albota2004, Roussev2004, Langrock2005,
Pelc2009}. We use $\eta_{ext}^M $ and $\eta_{n}$ as free
parameters. We find $\eta_{n} \sim 72 (7) \,\mathrm{\%/W \cdot
cm^2}$, which for our $3\,\mathrm{cm}$ long waveguide leads to a
total normalized conversion efficiency of $\eta_c = 650
(70)\,\mathrm{\%/W}$.

As can be seen in the plot, the maximum conversion efficiency is
achieved for $P_p \sim 400\,\mathrm{mW}$, corresponding to an
input power before the waveguide $P_{in}=1.33\,\mathrm{W}$, which
is far below the maximum output power of our amplifier ($\sim
5\,\mathrm{W}$).

The maximal internal conversion efficiency should be 1 in theory.
In practice, it is limited by internal waveguide losses. We have
not measured them directly, but according to the specification of
the manufacturer we can infer an effective transmission of
$\eta_{t} \sim 0.7$, assuming that photons are created at the
center of the waveguide. This would set the maximal achievable
value for $\eta_{int}^{M}$. The remaining discrepancy between the
inferred $\eta_{int}^{M}=0.41$ and $\eta_{t} \sim 0.7$ could be
explained by non perfect mode overlap and by variations of the
poling period along the waveguide \cite{Zaske2011}.

\subsection{Signal to noise ratio and pump power}
\label{sec:SNR} As explained in section \ref{sec:FWHM}, another
relevant figure of merit to analyze the conversion process is the
signal to noise ratio ($SNR$). Combining Eqs. (\ref{eq:N}) and
(\ref{eq:eta}), and recalling the definitions of $S$ and
$\eta_{\mathrm{tot}}$, we can derive the following relation:
\begin{equation}
    \label{eq:snr}
    SNR_{DC}(P_p) = \frac{p_S-p_N}{p_N} = \frac{\mu_{in} \cdot
    \eta_{\mathrm{tot}} \cdot f \left( P_p
    \right)}{\alpha \cdot P_p + DC}.
\end{equation}
The subscript $DC$ indicates that the signal to noise ratio is
taken without dark count subtraction, since here we want to
account for the limitations given by our detection system.

It has to be observed that the signal shows an oscillatory
behavior as a function of the pump power, while the noise
increases linearly. As a consequence, $SNR_{DC}$ initially
increases with $P_p$ until it reaches a maximum and then decreases
to zero. When dark counts are subtracted $SNR$ shows a maximum at
zero pump power and then decreases as $P_p$ increases. This
behavior can be seen in Fig. \ref{fig:SNR}(b), where the measured
$SNR_{DC}$ is plotted as a function of the pump power (blue dots,
right axis). The experimental data belongs to the same series
analyzed in the previous section. The blue continuous line is
obtained from Eq. (\ref{eq:snr}) with no free parameter. All the
values involved in the formula ($\eta_{\mathrm{tot}}$, $\alpha$,
{\it etc.}) have been independently measured. The blue shaded area
accounts for the experimental errors on these parameters. It is
remarkable that the expected theoretical curve reasonably agrees
with the measured data within the error bars.

As we observed in section \ref{sec:eff}, the pump power required
to achieve maximum efficiency is about $400\,\mathrm{mW}$. Under
this condition, the measured signal to noise ratio is only half of
its maximum value. On the other hand, at $P_p \sim
100\,\mathrm{mW}$ the signal to noise ratio reaches its optimum
for a conversion efficiency which is about half of
$\eta^M_{\mathrm{ext}}$. Since these two quantities cannot be
varied independently, the pump power has to be chosen in order to
find a tradeoff between efficiency and $SNR_{DC}$. For the
measurements described in the following sections, we choose $P_p
\sim 120\,\mathrm{mW}$.

\subsection{Signal to noise ratio and filtering bandwidth}
\label{sec:BW}As detailed in section \ref{sec:N}, the noise level
detected in the experiment is linearly proportional to the pump
power. We noticed that $N$ cannot be significantly reduced by
introducing an extra band-pass filter centered at
$1552\,\mathrm{nm}$, with a transmission of only $10^{-5}$ at
$1569\,\mathrm{nm}$. This observation indicates that the noise
level is only weakly related to leakage of the pump, and suggests
that it could be mainly due to spontaneous Raman scattering of the
$1569\,\mathrm{nm}$ light. The spectrum of Raman scattering
typically extends over a region of several hundreds of nanometers,
and it could then have a significant contribution around the
converted wavelength $\lambda_{out}$. If this is the case, we
should observe a variation of the noise level related to a change
in the filtering bandwidth $\Delta\lambda$.

\begin{figure}[ht!]
\begin{center}
\begin{tabular}{cc}
\multicolumn{2}{c}{\includegraphics[width=0.9\textwidth]{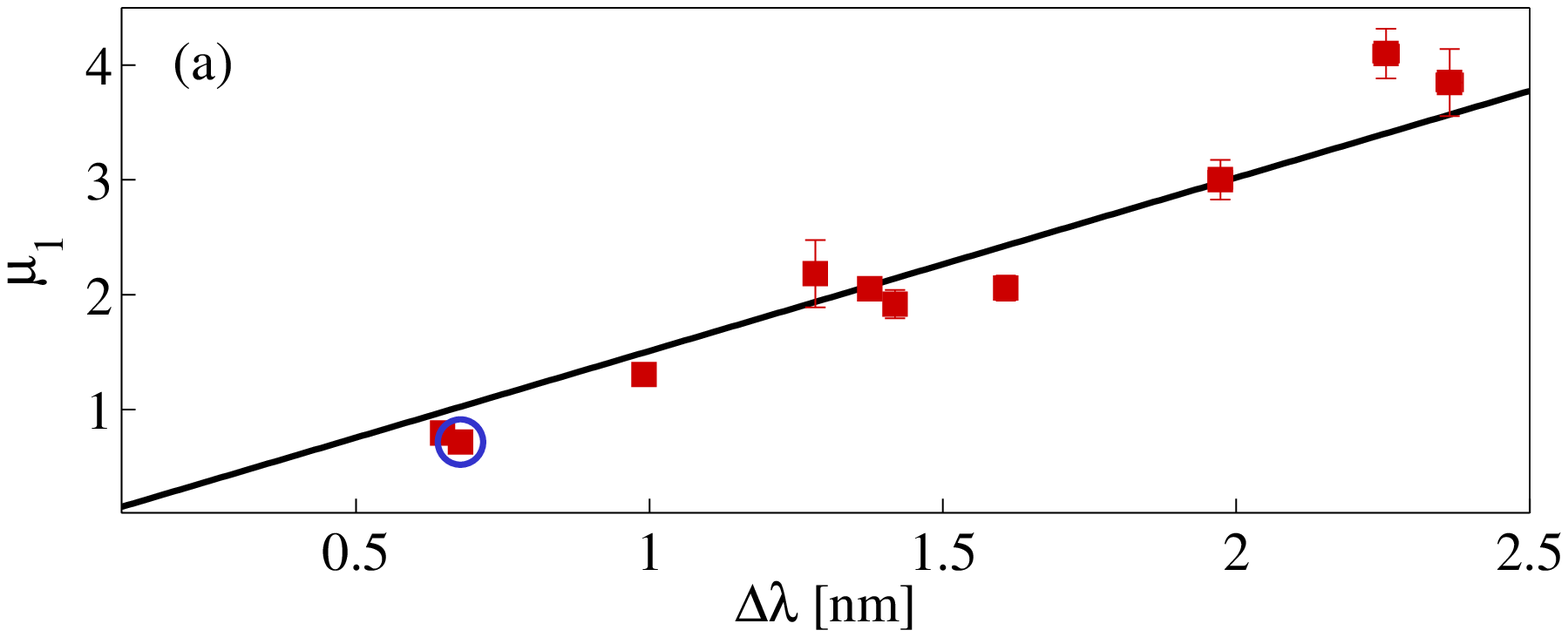}}
\\
\includegraphics[width=0.45\textwidth]{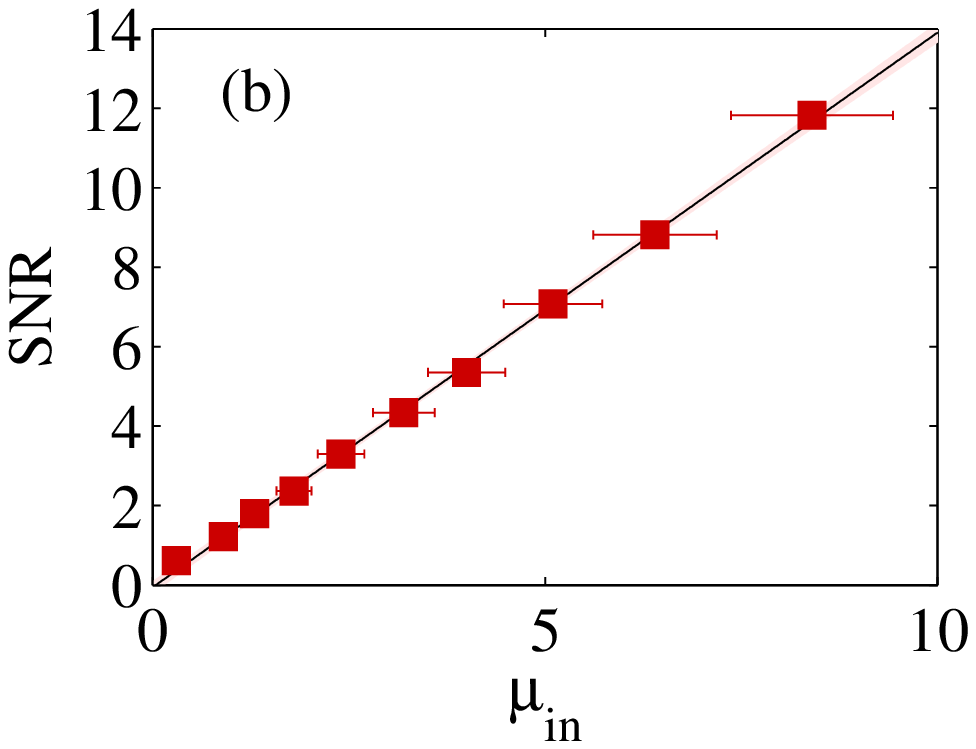} &
\includegraphics[width=0.45\textwidth]{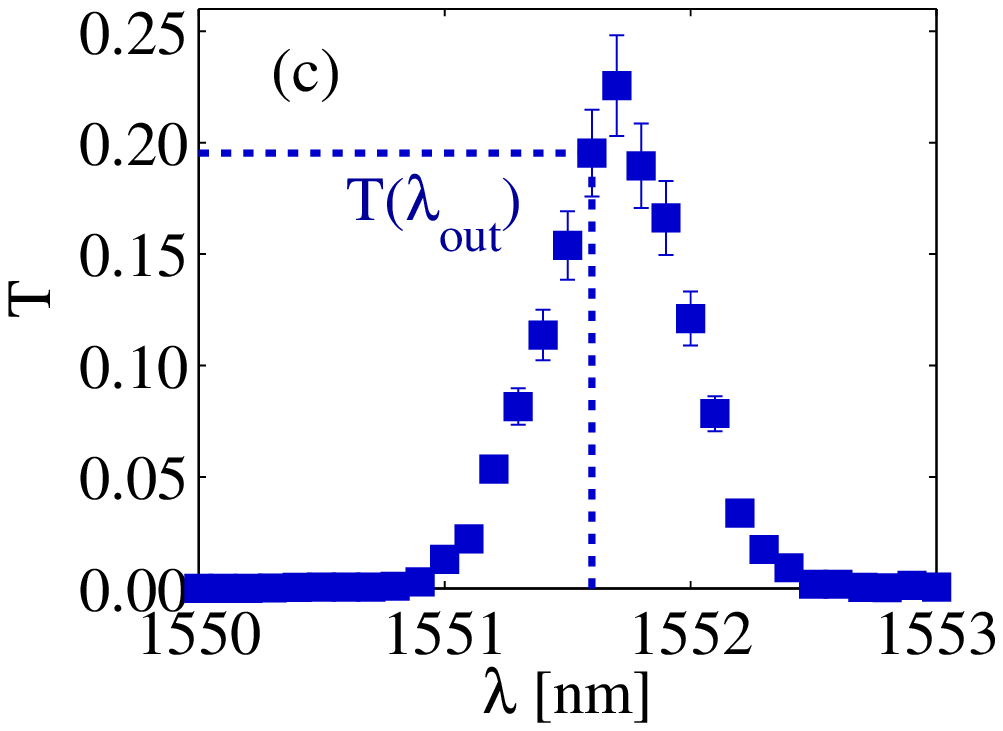}
\end{tabular}
\caption{\label{fig:m1}{\bf (a)} The mean number of photons in the
input pulse required to achieve a SNR = $1$ ($\mu_{1}$) is plotted
as a function of the filter width ($\Delta\lambda$). The solid
line is a linear fit with no offset. {\bf (b)} Signal to noise
ratio as a function of the mean input photon number. The data
shown correspond to the point highlighted with a blue circle in
(a). In this case $\mu_{1} = 0.7\pm0.1$. {\bf (c)} Transmission of
the filtering stage as a function of wavelength for the data
points shown in (b). Here the linewidth is
$\Delta\lambda=0.68\pm0.01\,\mathrm{nm}$. The vertical dotted line
indicates the target wavelength
$\lambda_{out}=1551.6\,\mathrm{nm}$. In this case the filter
transmission at $\lambda_{out}$ is $T (\lambda_{out}) =
0.23\pm0.01$.}
\end{center}
\end{figure}

In order to confirm this intuition, we proceed as follows. For a
given value of $\Delta\lambda$ we measure the signal to noise
ratio ($SNR$, with dark counts subtraction) for the converted
photons as a function of the mean input photon number $\mu_{in}$
(see Fig. \ref{fig:m1}(b)). We measure then the quantity $\mu_1$,
defined as the value of $\mu_{in}$ required to get $SNR=1$. We
repeat this measurement for different values of the filtering
bandwidth and observe the dependence of $\mu_1$ as a function of
$\Delta\lambda$. Assuming a constant level of Raman noise in the
vicinity of $\lambda_{out}$, the noise level is expected to be
proportional to the bandwidth of the filtering stage.

As we observed in section \ref{sec:filter}, the filtering
bandwidth can be varied by moving the out-coupling lens L2 along
its axis. For each position of L2, we measure the transmission of
the filtering stage as a function of wavelength which allows us to
determine the filtering bandwidth $\Delta\lambda$. To perform this
measurement, we replace the pump laser with a tunable laser source
(Tunics Plus, from Photonetics) and scan its wavelength while
monitoring the transmitted power through the filtering stage. A
typical curve is shown in Fig. \ref{fig:m1}(c). Note that, since
the system is aligned to maximize the total transmission, it may
happen that the maximal transmission of the grating $T_{max}$ is
at a wavelength slightly different from $\lambda_{out}$. However,
the transmission at $\lambda_{out}$ stands always above $90\,\%$
of $T_{max}$.

In Fig. \ref{fig:m1}(a) we plot $\mu_1$ as a function of
$\Delta\lambda$. The solid line is a linear fit with no offset.
The data points are in good agreement with the linear fit, which
confirms that the noise is proportional to the filtering width, at
least in this bandwidth range. For $\Delta \lambda
=0.68\,\mathrm{nm}$ ($\sim85\,\mathrm{GHz}$), we observe $\mu_1
\sim 0.7 < 1$. This measurement allows us to conclude that the
main source of noise for our setup has a broadband nature in the
spectral region explored in the experiment. This suggests that the
noise level could be significantly reduced by using a narrower
filter (see section \ref{sec:dis}). Similar results have been
recently achieved by Kuo et al. \cite{Kuo2013}. Together with
previous observations, this also suggests that the noise is
induced by spontaneous Raman scattering of the pump light.
However, in order to unambiguously confirm that this is the case,
a more detailed study would be required (see
\cite{Langrock2005,Zaske2011,Pelc2011}).

\section{Coherence preservation}
\label{sec:coh} In order to be used as a quantum interface, it is
important to demonstrate not only that the device operates in the
low noise regime, but also that the conversion is done in a
coherent fashion \cite{Ding2010} and preserves the quantum
superposition of an incoming qubit \cite{Curtz2010, Takesue2010}.
To test the preservation of quantum coherence we perform an
interference experiment with a pair of input pulses with a time
difference of $\tau = 100\,\mathrm{ns}$, and a phase difference
$\phi$. This state can be seen as a time-bin qubit and can be
written as $|\psi\rangle_{in}=|e\rangle+ e^{i\phi}|l\rangle$ where
$|e\rangle$ and $|l\rangle$ represent the early and late
time-bins, respectively.

A fiber based interferometer is placed before the single photon
detector in order to measure the coherence of the time-bin qubit
(see Fig. \ref{fig:setup}). The interferometer is realized
combining two $50$-$50$ fiber beam splitters. In order to adjust
the relative polarization between the two paths, a polarization
controller is inserted in the short arm, while a $20\,\mathrm{m}$
single mode fiber in the long arm ensures a time delay of
$\approx100\,\mathrm{ns}$ between the two paths. The relative path
length is not actively stabilized, but the interferometer is kept
inside a polystyrene box to passively improve its temperature
stability. The visibility of the interference fringes for
classical light at $1569\,\mathrm{nm}$ is measured to be
$\sim96\,\%$, limited in part by the seed laser linewidth of
$500\,\mathrm{kHz}$. Note that, when performing interference with
the converted photons, the linewidth of the input laser should
also be taken into account.

If the travel time difference between the two arms of the
interferometer corresponds to the time difference between the two
input pulses $\tau$, the pulses can exit the interferometers in
three different time slots. The first and the last times slots
corresponds to events $|e,S\rangle$ and $|l,L\rangle$,
respectively,  where $S$ and $L$ denote the short and long arms of
the interferometer. The central time corresponds to two
indistinguishable processes that can interfere: $|e,L\rangle$ and
$|l,S\rangle$. The count rate in this time slot displays an
interference as a function of the relative phase ($\phi-\gamma$),
where $\gamma$ is the phase introduced by the interferometer. The
visibility $V$ of this interference fringe is a measure of the
coherence preservation. To measure the visibility, we record
counts in the central time slots as a function of the phase
$\gamma$, for a pump power of $130\,\mathrm{mW}$.

\begin{figure}[htbp]
\begin{center}
\begin{tabular}{cc}
\multicolumn{2}{c}{\includegraphics[width=0.9\textwidth]{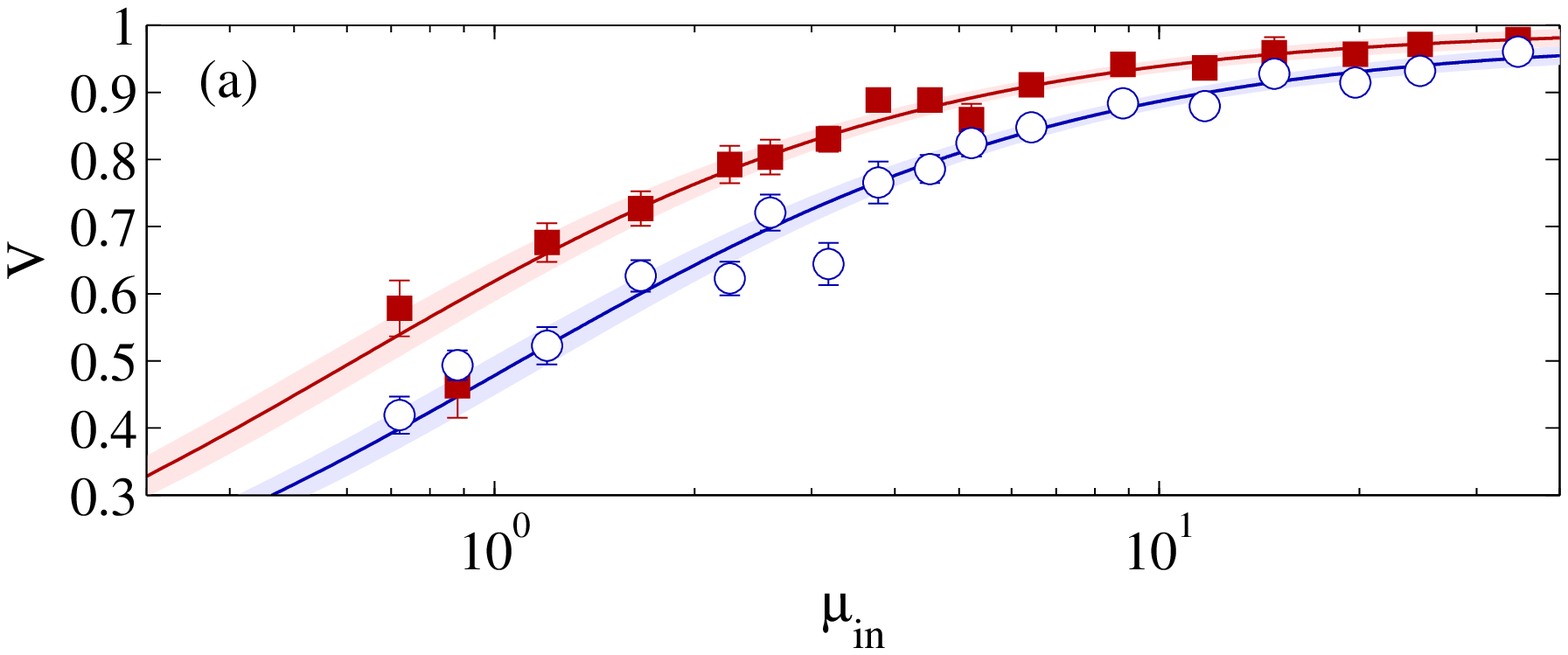}}
\\
\includegraphics[width=0.45\textwidth]{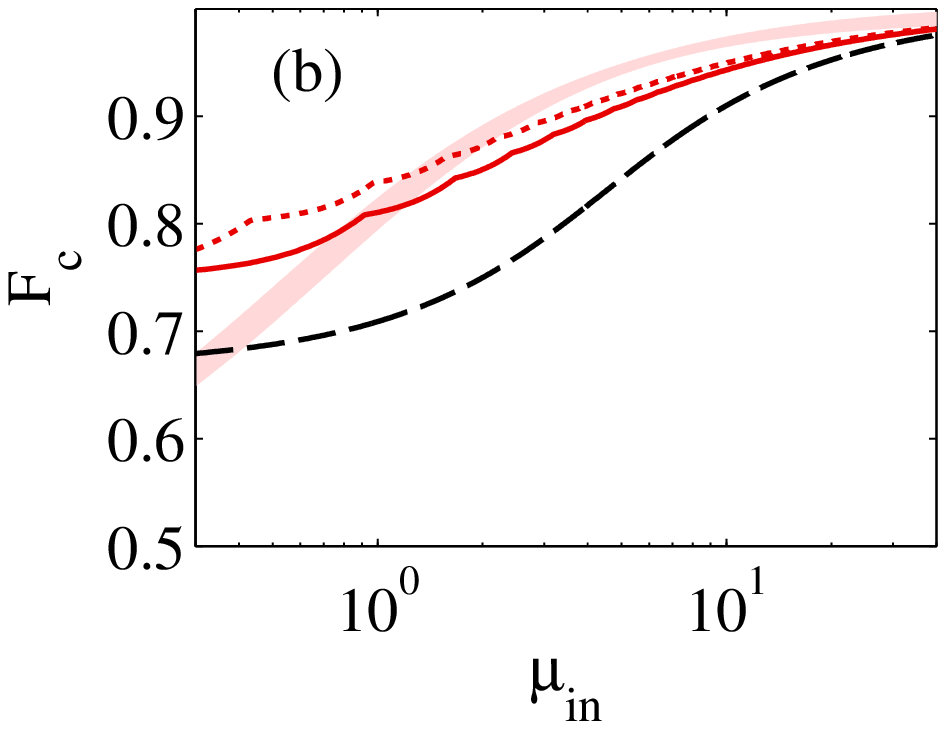} &
\includegraphics[width=0.45\textwidth]{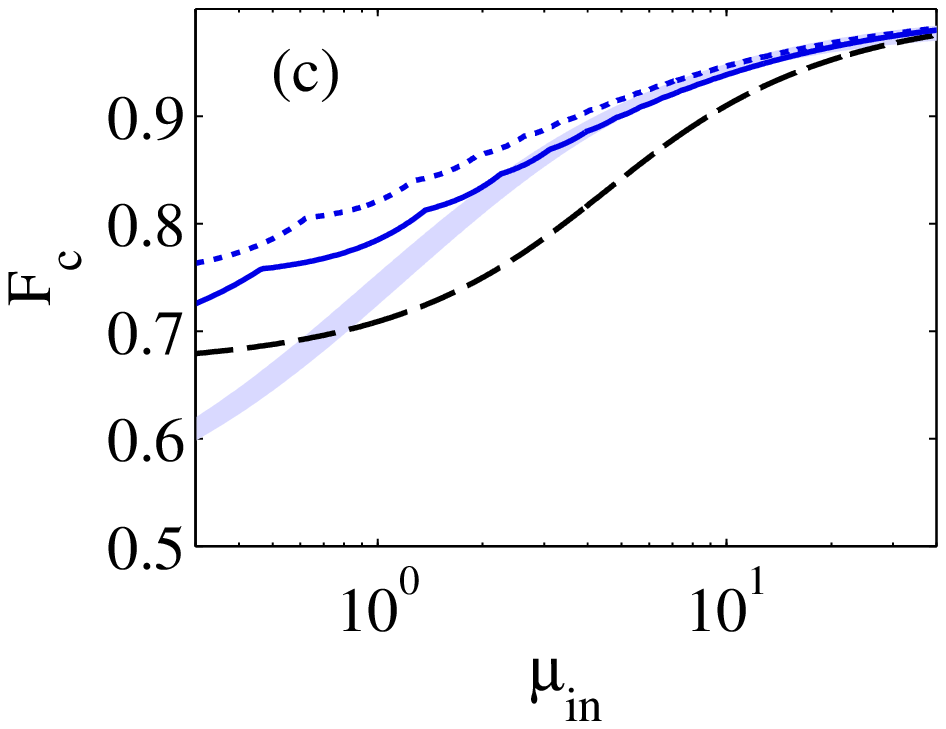}
\end{tabular}
\caption{\label{Fig_V}{\bf (a)} Visibility of the interference
fringes as a function of $\mu_{in}$ for $\Delta t_G =
20\,\mathrm{ns}$ (red plain squares) and $\Delta
t_G=50\,\mathrm{ns}$ (blue open circles). The pump power is
$130\,\mathrm{mW}$. The visibilities are corrected for the maximal
visibility of the interferometer $V_{max} =0.96$. The solid lines
are fits with Eq. (\ref{EqV}) and the shaded areas represent the
$95\%$ confidence interval of the fits. {\bf (b)} Comparison
between experimental fidelities measured for $\Delta
t_G=20\,\mathrm{ns}$ (shaded area taken from (a)) and the best
achievable fidelities using a classical {\it measure and prepare}
strategy. The dashed line corresponds to $\eta=1$, the solid line
to the external efficiency $\eta_{ext}=0.10$ and the dotted line
to device efficiency $\eta_{dev}=0.025$ inferred for these
measurements. {\bf (c)} Same as (b) but for $\Delta t_G
=50\,\mathrm{ns}$. The solid line corresponds to $\eta_{ext}=0.21$
and the dotted line to device efficiency $\eta_{dev}=0.055$.}
\end{center}
\end{figure}

Figure \ref{Fig_V}(a) shows a measurement of the visibility of the
interference fringes as a function of the average input photon
number per pulse $\mu_{in}$, for two different detection gate
widths, $\Delta t_G = 20\,\mathrm{ns}$ (plain squares) and $\Delta
t_G = 50\,\mathrm{ns}$ (open circles). Note that the visibilities
diplayed here are corrected for the maximal visibility of the
interferometer $V_{max}=0.96$. For high values of $\mu_{in}$, we
observe high visibilities $V>0.9$, suggesting that the conversion
process preserves the coherence to a great extent. However, the
visibility decreases when decreasing $\mu_{in}$ as a result of a
decrease of $SNR$. We also observe higher visibility for $\Delta
t_G = 20\,\mathrm{ns}$, which is consistent with the fact that the
decrease of visibility is due to higher noise level. To further
confirm this hypothesis, the data are fitted with a model taking
into account the $SNR$ to determine the visibility:
\begin{equation}
    V = V_0 \frac{\mu_{in}}{\mu_{in}+\mu_1/2}.
    \label{EqV}
\end{equation}
We observe an excellent agreement between this simple theoretical
model and the experimental data.

Even if the light used to characterize the phase preservation is
classical, it has been shown that it is possible to obtain
information about the quantum character of the conversion
\cite{Specht2011,Gundogan2012}. The idea is to measure the
conditional (i.e. conditioned on a successful conversion of the
qubit) fidelity $F_c$ of the output qubit with respect to the
input qubit and the compare the values obtained experimentally
with the best obtainable fidelity using classical strategies. The
conditional fidelity is defined as $F_c =
\langle\psi_{in}|\rho_{out}|\psi_{in}\rangle$, where $\rho_{out}$
is the density matrix of the output qubit and $|\psi_{in}\rangle$
is the state of the input qubit. It has been shown that if the
qubits are encoded in Fock states, the classical strategy that
maximize the fidelity is a "measure and prepare strategy", where
the qubit is measured in some basis and a new qubit is prepared
according to the measurement result \cite{Massar1995}. For single
photon qubits, this gives a classical fidelity of 2/3. For qubits
encoded in weak coherent states, as it is the case in our
experiment, it is necessary to take into account the Poissonian
statistics of the input light field and the finite efficiency of
the quantum frequency conversion device, as shown in
\cite{Specht2011,Gundogan2012}. Under these conditions, the
maximal classical fidelity increases with the average input photon
number $\mu_{in}$.

For superposition states $F_c$ can be obtained from the visibility
$V$ from the relation $F_c=(1+V)/2$ \cite{Marcikic2003}. Note that
we should in principle take into account an average fidelity
including the fidelities for the basis states $|e\rangle$ and
$|l\rangle$. However, we assume here that these fidelities are
higher than the ones for superposition states, which is justified
by the fact that there is no coherence involved.  Figures
\ref{Fig_V}(b) and \ref{Fig_V}(c) show the comparison between the
best classical strategies and the experimental results (shaded
area) for $\Delta t_G = 20\,\mathrm{ns}$ and $\Delta t_G =
50\,\mathrm{ns}$, respectively. The different lines correspond to
different efficiencies. The dashed line corresponds to $\eta = 1$,
i.e. takes only into account the Poissonian statistics of the
input fields. The solid (dotted) line corresponds to the external
conversion (device) efficiency inferred for that experiment. For
$\Delta t_G = 20\,\mathrm{ns}$, we have $\eta_{ext} =0.11$ and
$\eta_{dev} =0.028$ while for $\Delta t_G = 50\,\mathrm{ns}$ , we
have $\eta_{ext} = 0.21$ and $\eta_{dev} =0.055$. The difference
between the values for $\Delta t_G  = 20\,\mathrm{ns}$ or
$50\,\mathrm{ns}$ is mostly due to the fact that in this case we
do not correct for the $\beta$ factor in the detection efficiency
during the backpropagation, since the noise is measured for a
given gate width. As can be seen in the figures, the best
classical fidelities increase when decreasing the efficiency, as
explained in \cite{Specht2011, Gundogan2012}.  We observe that for
$\Delta t_G = 20\,\mathrm{ns}$, the measured fidelities are higher
than the classical threshold  for $\mu_{in}$ between $2$ and $25$
in all cases. For $\Delta t_G  = 50\,\mathrm{ns}$, the device is
at the limit of operating in the quantum regime, when considering
$\eta_{ext}=0.21$.

\section{Discussion} \label{sec:dis}
We have shown that the noise can be reduced sufficiently to
operate the frequency converter with single photon level input
fields compatible with atomic quantum memories. However, in order
to use this device with single photon fields emitted by the atomic
memory, a significant reduction of noise is needed to compensate
for the finite retrieval efficiency of the memory and passive
losses between the memory and the converter. The measurement of
the $SNR$ vs filter bandwidth suggests that the noise is
broadband, and inversely proportional to the filter bandwidth.
Since our input photons are relatively long ($30\,\mathrm{ns}$
FWHM), the filter bandwidth could be ideally reduced down to $
\sim 50\,\mathrm{MHz}$, thus decreasing significantly the noise
and only weakly affecting the signal transmission (96 $\%$
transmission for a gaussian filter). In Section \ref{sec:N}, we
have inferred an unconditional noise floor after the crystal of
$\alpha^\prime = 5\times10^{-6}\,\mathrm{mW}^{-1}\mathrm{ns}^{-1}$
for a filter linewidth of $85\,\mathrm{GHz}$. From this value, and
assuming a linear scaling down to the $\mathrm{MHz}$ level, we
infer that for a $50\,\mathrm{MHz}$ bandwidth we should obtain
$\alpha^\prime \sim 3\times 10^{-9} \mathrm{mW}^{-1}
\mathrm{ns}^{-1}$. This would lead to an unconditional noise floor
of $6 \times 10^{-5}$ photons at the maximum conversion efficiency
and for $\Delta t_G =50\,\mathrm{ns}$. This number is only given
to infer the final limitation of our device. Further measurements
should be performed to confirm the validity of the linear behavior
to much narrower bandwidth. In order to reach this narrow
bandwidth, phase shifted Fiber Bragg grating \cite{Kaiser2013} or
Fabry-Perot cavity could be used. It is likely that such narrow
filter would have to be actively stabilized with respect to the
converted light frequency. This may result in significant increase
of the technical complexity of the setup and it could potentially
introduce additional noise in the system. However, recent results
on the experimental realization of medium finesse ($\sim200$)
monolithic resonators \cite{Palittapongarnpim2012} suggest that a
narrower filtering could be achieved without a substantial
increase of the setup complexity. A significant noise reduction
could already be achieved with a fiber Bragg grating filter with
$1$ -- $2\,\mathrm{GHz}$ bandwidth and passive thermal stability,
leading to an inferred $\mu_1 <0.02$. This should be sufficient to
obtain high $SNR$ with single photon input and realistic memory
retrieval efficiencies.

\section{Conclusion}
We have presented a photonic quantum interface capable of
converting weak atomic resonant light to telecommunication
wavelength. We showed that weak coherent states with $\mu_{in} <
1$ with a duration of $30\,\mathrm{ns}$ can be converted from
$780\,\mathrm{nm}$ to $1552\,\mathrm{nm}$ and detected with a
signal to noise ratio above $1$. We showed that the conversion
process preserves the coherence of time-bin input qubits and that
the device can operate in the quantum regime. We revealed the
broadband nature of our main source of noise and have measured an
unconditional noise floor after the crystal of $5 \times 10^{-6}$
photons per $\mathrm{mW}$ of pump power and per nanosecond, for a
filter spectral bandwidth of $85\,\mathrm{GHz}$. In addition, we
showed that the noise level can be controlled tuning the bandwidth
of the filtering stage. This suggests that much lower noise level
can be reached using narrower filters. This could lead to an
unconditional noise floor sufficiently low to enable practical
frequency conversion of non classical light emitted from atomic
quantum memories using an integrated waveguide device.

\section*{Acknowledgments}
We acknowledge financial support by the  ERC starting grant QuLIMA
and by the Spanish MINECO (project FIS2012-37569). We thank
Valerio Pruneri, Davide Janner and Vittoria Finazzi for lending us
the tunable laser at 1550 nm and for stimulating discussions.

\end{document}